\documentclass[11pt,twoside]{article}

\usepackage{asp2006}
\usepackage{epsf}
\usepackage{lscape}

\begin{document}
\title{Non-Thermal Radio Emission from \\ Colliding-Wind Binaries}   
\author{Ronny Blomme}   
\affil{Royal Observatory of Belgium, Ringlaan 3, B-1180 Brussel, Belgium}    

\begin{abstract} 
In colliding-wind binaries, shocks accelerate a fraction of the electrons
up to relativistic speeds. These electrons then emit synchrotron radiation
at radio wavelengths. Whether or not we detect this radiation depends on the
size of the free-free absorption region in the stellar winds of both
components.  One expects long-period  binaries to be detectable, but not the
short-period ones. It was therefore surprising to find that Cyg OB2 No. 8A
(P = 21.9 d) does show variability locked with orbital phase.

To investigate this, we developed a model for the relativistic electron
generation (including cooling and advection) and the radiative transfer
of the synchrotron emission through the stellar wind. Using this model,
we show that the synchrotron emitting region in Cyg OB2 No. 8A does
extend far enough beyond the free-free absorption region to
generate orbit-locked variability in the radio flux.

This model can also be applied to other non-thermal emitters and will prove
useful in interpreting observations from future surveys, such as COBRaS --
the Cyg OB2 Radio Survey.
\end{abstract}



\section{Introduction}

For most hot stars, the radio radiation they emit is due to 
thermal free-free emission by the
ionized material in the stellar wind. A number of stars, however,
also show \emph{non-thermal} radio 
emission, characterized by a negative spectral 
index\footnote{The spectral index is given by $\alpha$, the exponent in
     the power law relation between flux and frequency: 
     $F_\nu \propto \nu^\alpha$. A stellar wind with a thermal spectrum
     has $\alpha \approx +0.6$.}.
This non-thermal emission is believed to be due to electrons that
are Fermi-accelerated around shocks \citep{RB_Bell78}.
As these relativistic electrons spiral around in the magnetic
field, they
emit synchrotron radiation, which we detect as non-thermal radio
emission \citep{RB_Bieging+al89}. In recent years, it has become
clear that the shocks responsible for the Fermi-acceleration in massive
stars are
those created in the colliding-wind region of a binary
\citep[][Van Loo, these proceedings]{RB_VanLoo+al06}.

Non-thermal radio emission from massive stars in general has
been reviewed by Benaglia (these proceedings).
In the present paper we will be mainly concerned with one 
specific colliding-wind binary,
Cyg~OB2 No.~8A. From spectroscopic observations,
\citet{RB_DeBecker+al04,RB_DeBecker+al06} 
found it to be an O6If + O5.5III(f)) system
with a period of
21.908 $\pm$ 0.040 d and an eccentricity of 0.24 $\pm$ 0.04.
Other parameters relevant for our work 
are listed in Table~\ref{table parameters}.

\begin{table}
\caption{Parameters of Cyg OB2 No. 8A used in this work.
Data from \citet{RB_DeBecker+al06}.}
\label{table parameters}
\begin{center}
\begin{tabular}{llllll}
\tableline
\noalign{\smallskip}
 & primary & secondary \\
\noalign{\smallskip}
\tableline
\noalign{\smallskip}
$T_{\rm eff}$ (K)                & 36800 & 39200 \\
$R_*$ ($R_{\sun}$)                 & 20.0  & 14.8 \\
$M_*$ ($M_{\sun}$)                 & 44.1  & 37.4 \\
log $L_{\rm bol}/L_{\sun}$         & 5.82  & 5.67 \\
$\dot{M}$ ($M_{\sun} {\rm yr}^{-1}$) & $4.8 \times 10^{-6}$ & $3.0 \times 10^{-6}$ \\
$v_\infty$ (km s$^{-1}$)           & 1873  & 2107  \\
\noalign{\smallskip}
\tableline
\end{tabular}
\end{center}
\end{table}

This binary has also been known
to be a non-thermal radio emitter since the first major survey
of O-type star radio emission by \citet{RB_Bieging+al89}.
It was recognized as such by its clearly negative spectral index.
The colliding-wind nature of this object is further confirmed by information
from its X-ray emission.
Observations by \citet{RB_DeBecker+al06} show an essentially thermal 
X-ray spectrum, 
but with an overluminosity of a factor 19-28 compared to the canonical 
$L_X/L_{\rm bol} = 10^{-7}$
ratio for O-type stars. This overluminosity is due to additional
X-ray emission by material heated
in the colliding-wind region. The X-ray light curve also
shows variability at the level of $\sim$~20 \%. Although the light curve is not 
well sampled,
it suggests that the variability is phase-locked.

Colliding-wind systems such as Cyg~OB2 No.~8A are important 
objects because they can provide information on
the Fermi acceleration mechanism, which is also relevant to other
fields of astrophysics, such as 
interplanetary shocks and supernova remnants.
The colliding-wind systems can also be used to constrain the amount of
clumping and porosity in stellar winds. This is important for
mass-loss rate determinations in single stars, which are
subject to considerable uncertainties \citep{RB_Puls+al08}.
Finally, the non-thermal radio emitters 
can help us detect binaries which are difficult to find spectroscopically
and they are therefore useful
for improving the determination of the binary frequency in clusters.

\section{Radio Variability}

\begin{figure}[!ht]
\plotone{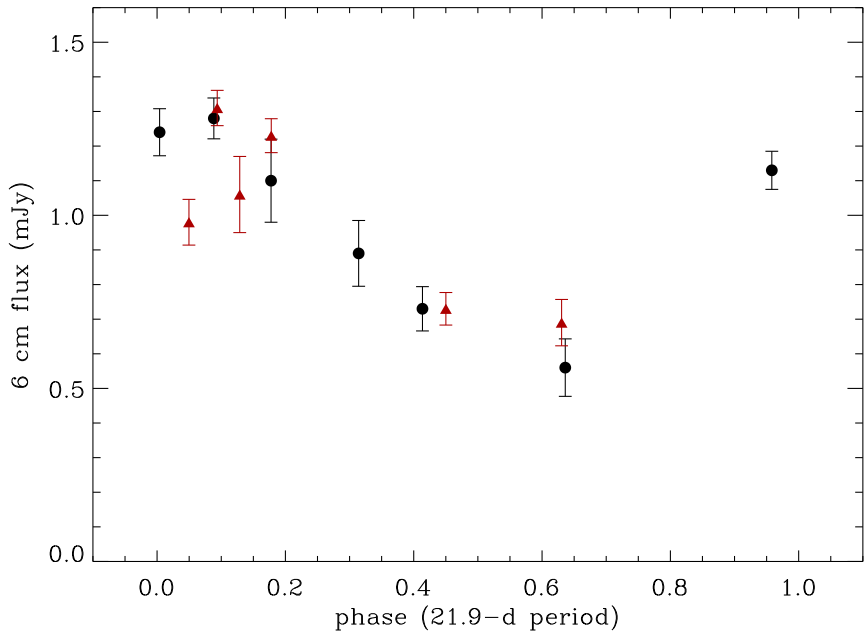}
\caption{6~cm radio fluxes of Cyg OB2 No.~8A, as a function of 
   orbital phase. The filled circles indicates data 
   from a first orbit, the filled triangles data from the subsequent orbit.}
\label{figure fluxes}
\end{figure}

Because of the importance of the colliding-wind binary Cyg~OB2 No.~8A,
we decided to monitor its radio flux.
We obtained a number of 6~cm continuum observations with the
NRAO\footnote{The National Radio Astronomy
             Observatory is a facility of the National Science Foundation
             operated under cooperative agreement by Associated Universities,
             Inc.}
Very Large Array (VLA) during 2005 February~04 to March 12, covering about 1.6
orbital periods.
Figure~\ref{figure fluxes} plots these data as a function of phase in the
orbital period. The radio fluxes show clear variability and the variability
is phase-locked with the orbital period. Furthermore, the data repeat
well from one orbit to the next. Since these data were collected, we
have also reduced VLA archive data (Blomme et al. 2009, in preparation).
These confirm the phase-locked variability and the orbit to orbit repeatability
seen in Fig.~\ref{figure fluxes}.

It is, however, quite surprising that phase-locked variability is found
in such a short-period binary. The main part of the colliding-wind 
region is expected to be comparable in dimension to the separation between
the two stars. This size (of order a few stellar radii) is quite small 
compared to the size of the free-free absorption region of each of the
components (of order 100 stellar radii). We therefore expected all the
synchrotron emission from the colliding-wind region to be absorbed by the
free-free absorption. Hence non-thermal radio emission 
from Cyg~OB2 No.~8A should have
been undetectable.

\citet{RB_Blomme05} made a simple model for this, assuming the synchrotron
emission to be a point source positioned between the two stellar components. 
Following the radiative transfer of the synchrotron emission through
the free-free absorption region showed, as expected, no detectable
level of non-thermal emission. As the data clearly contradict this, 
the effect of porosity was also investigated. The porosity variant of
the model assumes that
the wind is clumped and that at least part of these clumps are optically
thick. This allows radiation to escape more easily through the holes
between the clumps, allowing us to look deeper into the stellar wind. 
Porosity models were indeed marginally capable of
explaining the fact that we do detect non-thermal radiation from
Cyg~OB2 No.~8A.
This conclusion is important, also in the context of single stars, as
it would help explain the important discrepancies found in the mass-loss
rate determinations of massive stars \citep[e.g.][]{RB_Puls+al08}.

\section{Modelling}

The point-source approximation for the synchrotron emitting region
is certain to be too simple.
We here introduce
a more sophisticated model for the colliding-wind region and its
associated radio emission.
The model does not solve the equations of hydrodynamics.
Instead it defines the geometric position of the contact discontinuity between 
the two winds using the equations from \citet{RB_Antokhin+al04}. On either
side of the contact discontinuity a shock is formed. To simplify
the model, we assume that these shocks and the contact discontinuity are
at the same geometric place. 

At the shocks new relativistic electrons are generated due to the
Fermi-acceleration mechanism. Their momenta follow a power-law distribution
with a high-energy cut-off due to inverse Compton
cooling. The fraction of electrons that become relativistic is determined
by assuming that 5\% of the shock energy goes into the electron acceleration.
We then follow the electrons as they are advected outward along the
contact discontinuity. We take into account that they cool down as
they do so (due to inverse Compton and adiabatic cooling) and we follow
them till they no longer emit synchrotron radiation
at the radio wavelengths we are interested in.

Next, we calculate the synchrotron emissivity along the contact discontinuity.
We do this in a 2-dimensional model. As the situation is rotationally
symmetric around the line connecting the two components, we can then
rotate this
2-dimensional emissivity into a 3-dimensional simulation box. Free-free
absorption and emission are also added to the model. Finally,
we determine the emergent intensity and flux using the
radiative transfer technique developed by 
\citet{RB_Adam90}.
We calculate this for a number of phases along the orbit, thus determining
the flux variability with orbital phase.
The input parameters we use in this model are listed in
Table~\ref{table parameters}. We also assumed an inclination angle of
32\deg, as estimated by \citet{RB_DeBecker+al06}.

\begin{figure}[!ht]
\plotone{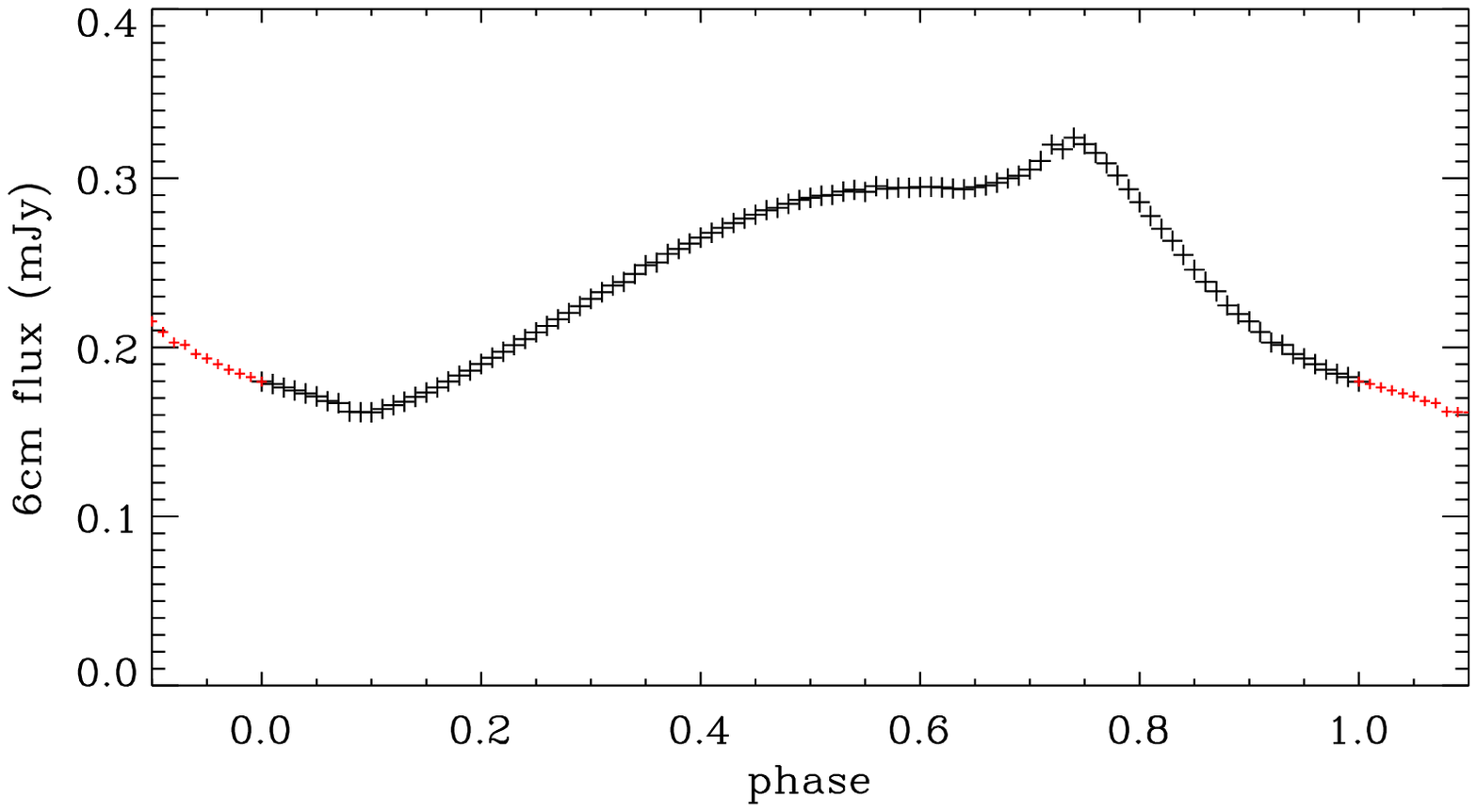}
\plotone{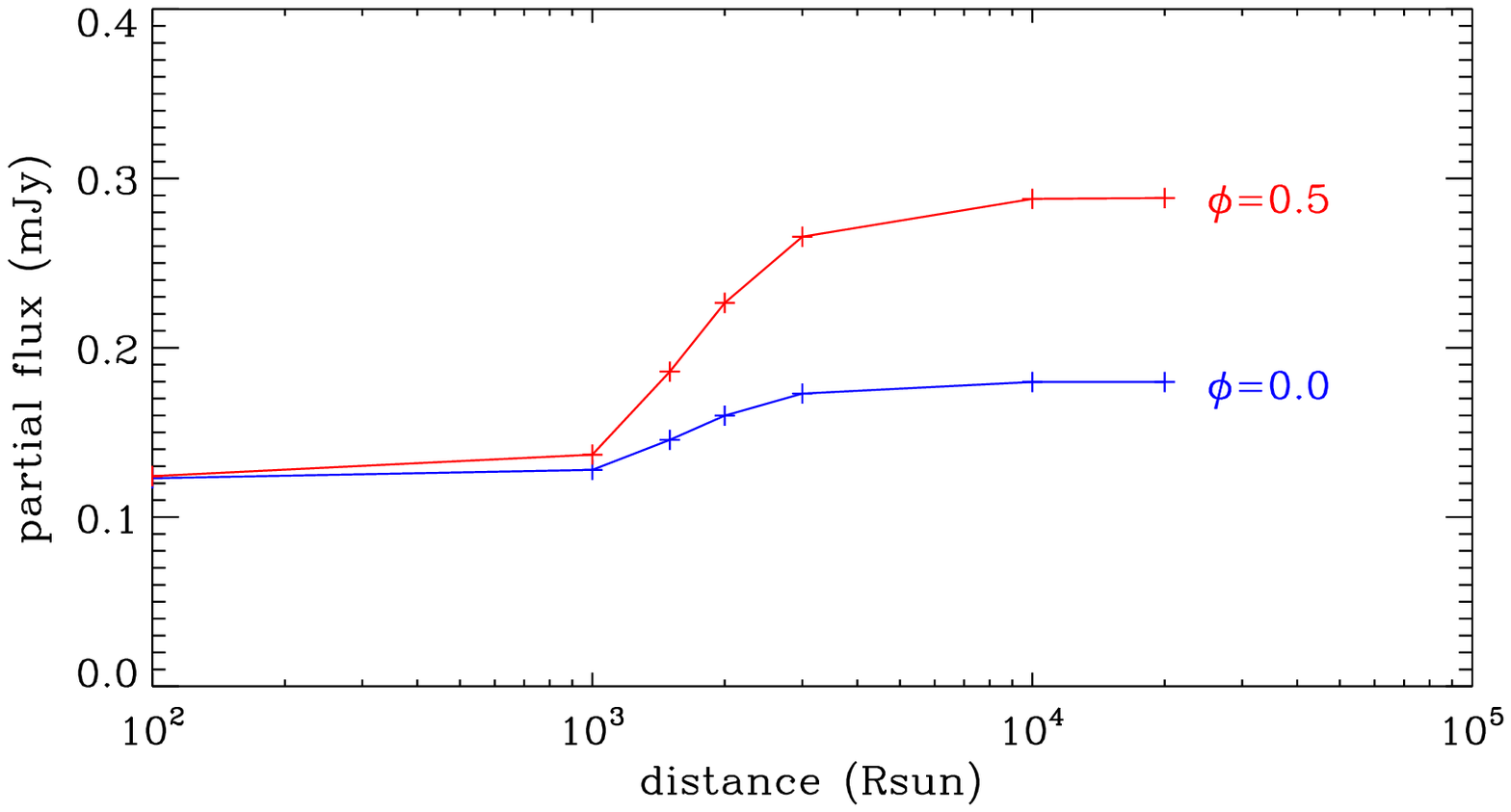}
\caption{Results from the more sophisticated synchrotron emission
  model. {\bf Top panel:} the 6~cm
  radio fluxes predicted by the model, as a function of orbital phase.
  {\bf Bottom panel:} the cumulative flux at phases 0.0 and 0.5, as a function
  of distance from the central line connecting the two stars. The
  largest contribution of the flux occurs between 1000--3000 $R_{\sun}$,
  i.e. approximately 10 to 30 times the separation between the two stars.}
\label{figure model 1}
\end{figure}

The top panel of Fig.~\ref{figure model 1} 
shows the resulting 6~cm radio fluxes.
Clearly, this more sophisticated model is capable of
providing variable non-thermal radiation, showing that at least part of the
synchrotron emission can escape from the free-free absorption. 
The observed level of non-thermal radiation (about 1 mJy max) is not
reached, but that could be due to one of the many assumptions
in the model, such as the fraction of the shock energy that 
goes into the relativistic electrons or the magnitude of the magnetic field.

The reason that this model does predict non-thermal radiation is that
the synchrotron emission region is considerably
larger than expected. This may seem surprising, but it is due to the
fact that at larger distances from the central line connecting the two stars
a much larger volume contributes to the synchrotron emission (as the
situation is rotationally symmetric around that line). That partly
compensates for the fact that the synchrotron emission per volume
becomes weaker as we move away from the central line. The model
also predicts that
in high-resolution radio images where we can resolve the colliding-wind
region we will still only see the central
part of the synchrotron emission region, as the emergent intensity
of the outer part will  be
too low for it to be detected. In this way, the model is consistent with 
the high-resolution radio observations 
\citep[e.g. WR140,][]{RB_Dougherty+al05}.

We next try to quantify the extent of the non-thermal emitting region,
by determining where the major
part of the detected non-thermal flux comes from. We do this by numerical
experiment: we artificially set the synchrotron emission to zero beyond
a certain radius and see how much non-thermal radiation is still
detected. The bottom part of Fig.~\ref{figure model 1} shows that
with a small cut-off radius we only obtain about 0.12~mJy, which is the
free-free emission of the two stellar winds. In the region 
1000--3000 $R_{\sun}$ we get the main contribution, with the flux
levelling off beyond that. The major flux contribution therefore
comes from distances of 1000--3000 $R_{\sun}$,
i.e. approximately 10 to 30 times the separation between the two stars.

Although the new model is quite successful in predicting non-thermal
radiation, there is still one major problem: the phases at which
maximum and minimum occur are incorrect. Observed maximum is around phase
0.1 and minimum at 0.7, but the predicted values are nearly exactly the
reverse.

\begin{figure}[!ht]
\plotone{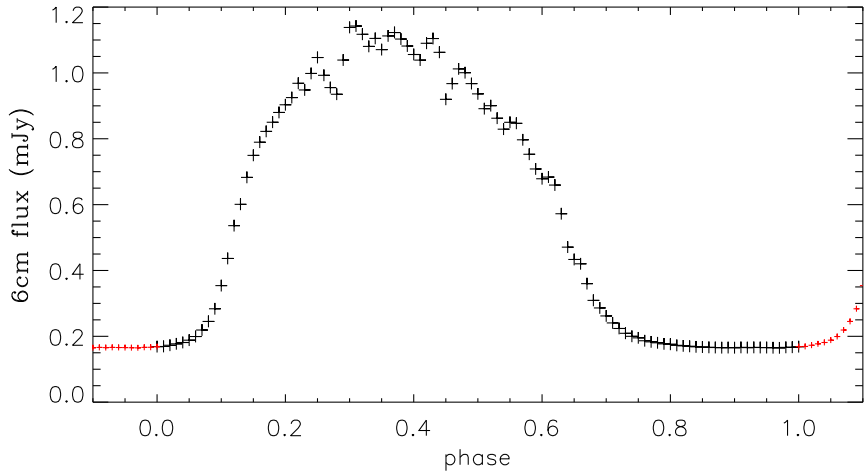}
\caption{Predicted 6~cm radio fluxes as a function of orbital phase
for a model with changed stellar parameters compared to those listed
in Table~\ref{table parameters}. Here we use $v_{\infty,1} = 3500$~km\,s$^{-1}$,
$\dot{M}_1=2 \times 10^{-6}$ M$_{\sun}$\,yr$^{-1}$ and
$\dot{M}_2=6 \times 10^{-6}$ M$_{\sun}$\,yr$^{-1}$.}
\label{figure model 2}
\end{figure}

The parameters listed in Table~\ref{table parameters} are based on typical
values for the spectral types of both components
\citep{RB_DeBecker+al06}. Quite an error range is associated
with these parameters, and we therefore explored if we could get
a better fit to the observed data by changing some of the star and
wind parameters. No exhaustive search of the parameter space was attempted,
but in Fig.~\ref{figure model 2} we present an improved model,
where the terminal velocity of the primary and the mass-loss rate
of the secondary are increased, while the mass-loss rate of the
primary is decreased. While
the phase is still not correct, the difference with the observations is less.
Furthermore,
the general flux level is also higher and in better accordance with the
observations.

To explain the remaining discrepancies, 
we explored a further sophistication of the model.
Following the work of \citet{RB_Parkin+Pittard08}, we introduced the
effect of orbital motion. In the simplification 
by \citeauthor{RB_Parkin+Pittard08}, this is done by assuming that the
contact discontinuity consists of a ``cap" in the inner part,
which is formed instantaneously and a ``ballistic" part in the outer part, 
where the
material streams out further without any net force
working on it. In the ballistic part, a ``time-delay" effect needs to
be taken into account: at any given phase, the contact discontinuity
in the ballistic part is the result of material emitted at the ends
of the cap at some previous phase.
Taking the orbital motion into account, this results in a spiral-shaped
form of the contact discontinuity in the orbital plane
\citep[see][their Fig. 10]{RB_Parkin+Pittard08}.

Although the preliminary calculations presented at the meeting
suggested that this would improve
the agreement with the observations, we have so far not succeeding
in finding a good model. As the
synchrotron emission region in this model is wound up into a spiral,
we detect a large
part of it at any orbital phase
and we therefore end up with considerably less variability as a
function of orbital phase. We suspect that 
in the present case of Cyg~OB2 No.~8A, 
the contact discontinuity is not as tightly wound up into
a spiral as suggested by the \citet{RB_Parkin+Pittard08} model.
Some twisting of the contact discontinuity would undoubtedly be present,
however, and this might be sufficient to explain the
phase shift between the observations and simpler models.

\section{Other Stars}

We intend to apply the present model to a number of other binaries
where we have a considerable amount of data on their
non-thermal radio emission, such as
HD 168112 \citep{RB_Blomme+al05}, 
HD 167971 \citep{RB_Blomme+al07}
and Cyg~OB2 No.~9 \citep{RB_VanLoo+al08}.
We will also extend the model to X-ray emission and optical spectra.

We also expect a considerable number of colliding-wind binaries to be
detected in future radio surveys. One upcoming survey is
COBRaS, the Cyg OB2 Radio 
Survey\footnote{{\tt http://www.homepages.ucl.ac.uk/$\sim$ucapdwi/cobras/}}
(PI: R.K. Prinja). This e-MERLIN 
legacy project has been awarded nearly 300 hrs to
make a deep survey of the core of the Cyg OB2 
cluster. This cluster has already provided us with three examples of
colliding-wind binaries (No.~5, No.~8A and No.~9).

In our Galaxy, there are only a few examples of young, very massive 
clusters, of which Cyg OB2 is one. The survey will make a mosaic
of the core region at 5 GHz (with a 1-sigma noise level of 3 $\mu$Jy) and 
at 1.6 GHz (1-sigma noise level of 7.5 $\mu$Jy). Its main science goals are
to study (i) the mass-loss and evolution in massive stars;
(ii) the formation, dynamics and content of massive OB associations;
and (iii) the frequency of massive binaries and the incidence of non-thermal 
radiation
The combination of the two frequencies 
will allow us to easily distinguish the non-thermal radio emitters from
the thermal ones and thereby detect the colliding-wind binaries.
The high sensitivity will result in the detection of 
a significant number of binary systems, allowing detailed statistical
studies of the colliding-wind phenomenon.

\section{Conclusions}

Contrary to expectations, the short-period binary Cyg OB2 No. 8A does
show radio variability that is phase-locked with its 21.9-day orbital period.
New models for the synchrotron emission from the colliding-wind region
show that this region is indeed extended enough that not all the emission
is absorbed by the free-free absorption. At present, we still have the problem 
that the phases of minimum and maximum flux are not correctly modelled. 
We suspect that
a combination of different wind parameters and some slight twist of
the contact discontinuity due to
orbital motion will be needed to explain these differences.

We note that
the present model does not require any porosity in the stellar winds
of both components. Colliding-wind systems such as Cyg OB2 No. 8A
can therefore be highly relevant to interpreting 
single stars, where clumping and porosity have been proposed to explain the
discrepancies in mass-loss rate determinations.

In future, the present models will be applied to other non-thermal
emitters. The model will also be extended to cover X-ray and optical
wavelengths. It is expected that future surveys, such as COBRaS, will
turn up a considerable number of non-thermally emitting colliding-wind
binaries.

\acknowledgements 
We thank Joan Vandekerckhove for his help with the reduction of the
VLA data.


\end{document}